\documentclass[aip,cha,reprint,groupedaddress]{revtex4-1}
\usepackage{graphicx}
\usepackage{bm}
\usepackage{mathtools}
\usepackage[normalem]{ulem}
\usepackage{comment}
\usepackage{caption}
\usepackage{hhline}
\usepackage{subcaption} 
\usepackage{natbib}
\usepackage{hyperref} 

\begin{document}
\title{Interplay of inertia and external forcing in Kuramoto model}
\author{Pratishtha Agnihotri and Sarika Jalan \email{Corresponding Author: sarikajalan9@gmail.com}}
\affiliation{Complex Systems Lab, Department of Physics, Indian Institute of Technology Indore, Khandwa Road, Simrol, Indore-453552, India}

\date{\today}
\begin{abstract}

The impact of external forcing is well studied in the Kuramoto model without inertia, but remains unclear for inertial Kuramoto oscillators (KMI) with bimodal intrinsic frequency distributions. This article fills that gap, showing that competition between external forcing and intrinsic bimodality can suppress the intermediate standing wave states of bimodal KMI by entraining oscillators to the external forcing. Using a self-consistent analytical framework, we show that, for a bimodal distribution, forcing makes the backward transition discontinuous, unlike the continuous transition in the unimodal case. Further, for a bi-delta distribution, we derive a closed form expression for the backward solution branch. These results clarify how intrinsic frequency structure shapes the effect of external forcing, with implications for biological systems (e.g., photoreceptor and pacemaker cells) and for pinning-control strategies in multi-agent networks.

\end{abstract}

\maketitle

\textbf{In the inertial Kuramoto model (KMI), a bimodal intrinsic frequency distribution produces more complex time-dependent behavior, such as standing and traveling wave states, because two distinct oscillator groups compete with each other. A bi-delta distribution offers a simple representation of this effect and is also relevant for real-world systems like power grids, where generators and consumers form separate frequency clusters. Moreover, in many realistic settings, an external forcing is present and acts as a reference frequency capable of entraining the oscillators. For unimodal distributions, this generally causes a smooth, continuous onset of synchronization. In contrast, bimodal systems exhibit a discontinuous transition. Through numerical simulations and a self-consistent analytical approach, we demonstrate that external forcing eliminates the standing wave state and produces a direct, discontinuous backward transition to incoherence, bypassing the usual intermediate standing wave regime.}

\paragraph*{\bf{Introduction:}}
Synchronization is a phenomenon observed in many natural and artificial systems, such as neural networks, circadian cycles, Huygens's clock, laser arrays \cite{Pele2006}, and networks of coupled electronic \cite{cosp2004synchronization} or mechanical oscillators \cite{pikovsky1985universal}. For a pair of oscillators, synchronization is often defined when they evolve with the same frequency and remain in the same phase over time due to mutual interactions. A group of oscillators with the same mean frequency forms a synchronized cluster. The Kuramoto model \cite{Kuramoto1975} provides a simple, widely used framework for studying how collective synchronization emerges from interactions within a population. Despite its simplicity, the model captures several essential features widely encountered in real-world systems, particularly transitions from incoherence to synchronized behavior and vice versa \cite{Bonilla}.
A second-order extension of the Kuramoto model was proposed by Tanaka \textit{et al} \cite{Tanaka1997_PD, Tanaka1997_PRL}, inspired by Ermentrout's modeling of synchronized flashing in the tropical Asian firefly \textit{Pteroptyx malaccae} \cite{Ermentrout1991}. This model has since been widely applied to systems such as power grids \cite{Ji2014, Filatrella2008, Rohden2012}, Josephson junction arrays \cite{Levi1978, WATANABE1994, Trees}, goods markets \cite{Ikeda2012}, and dendritic neurons \cite{SAKYTE20113912}. When we include inertia in the Kuramoto model, the system exhibits richer behavior, including abrupt synchronization transitions, hysteresis, and the coexistence of multiple stable states \cite{Simona_2014, ji2014low, ghosh2025universal}. These features make the inertial Kuramoto model better suited for describing many real-world systems.

Oscillators often interact with external driving signals, sometimes called leaders or pinners \cite{Inertia_forcing_2023}, which affect their collective behaviors. The interactions of second-order Kuramoto oscillators with external driving force have been widely studied in contexts such as photoreceptor cells in the eye \cite{Wang1992}, pacemaker via distributed control \cite{Li2015, Rao2018}, and pinning control in multi-agent systems \cite{Yang2013, Chu2017}. This forcing can significantly alter the synchronization process and give rise to additional dynamical states. Previous studies have examined the effects of external forcing and inertia in the Kuramoto model across various settings. Sakaguchi \cite{Sakaguchi_1988} demonstrated that external forcing in Kuramoto systems without inertia can give rise to either forced or mutual entrainment depending on the forcing and coupling strengths. Childs and Strogatz \cite{Strogatz_2008_forcing} provided one of the earliest systematic analysis of the resulting bifurcation structure, which Costa \cite{Bifurcation_HO_2024} later generalized to account for higher-order interactions.
On the other hand, studies on the inertial Kuramoto model without forcing have demonstrated the emergence of first-order synchronization transitions, chimera states, hysteresis, and bistability \cite{olmi2015chimera, Narayan_2024}. Some studies have examined the combined influence of inertia and forcing 
\cite{Inertia_forcing_2023}. 
However, these studies on inertia and external forcing have focused on unimodal frequency distributions, and a systematic understanding of how inertia and forcing jointly affect systems with bimodal distributions remains lacking. 

\begin{figure*}[t]
\includegraphics[width=1\textwidth]{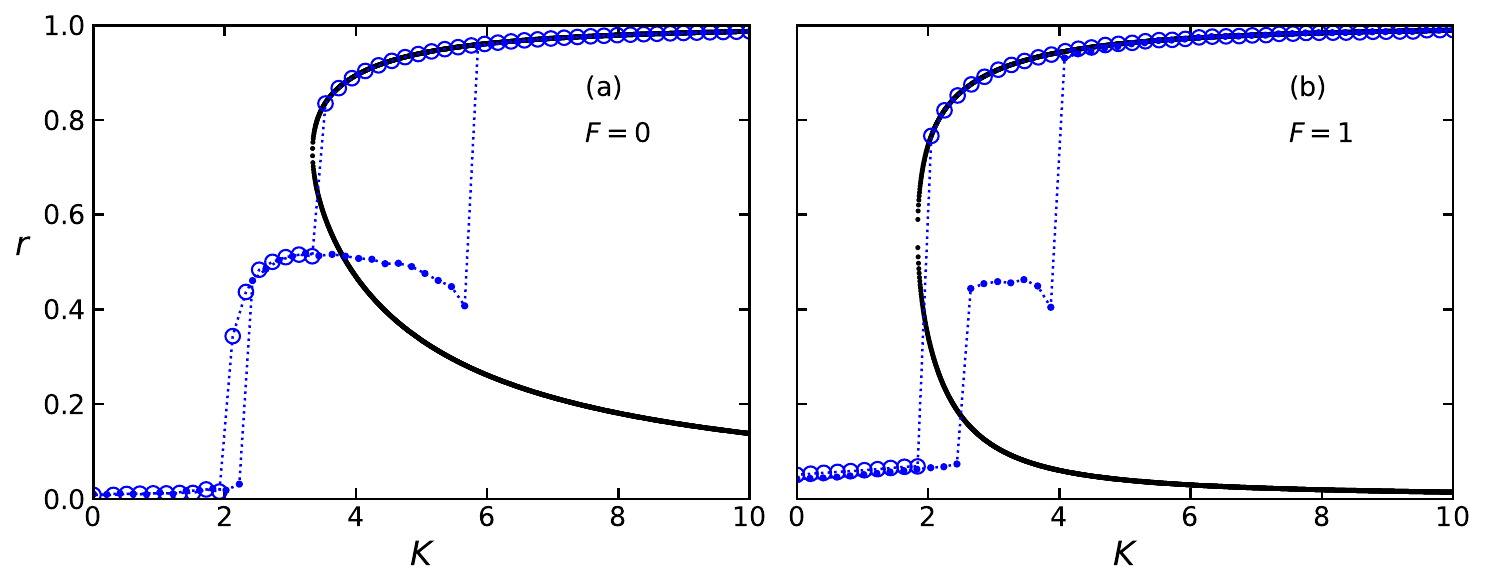}
\caption{
Order parameter $r$ as a function of coupling strength $K$ for a bimodal Gaussian frequency distribution with $m = 1$ and $N = 10^4$. Filled blue circles denote the forward branch, while open blue circles denote the backward branch; the black line represents the analytical solution for the backward branch obtained from Eq.~(\ref{eq10}).
}
\label{fig:rk_f0_f1}
\end{figure*}

The distribution of intrinsic frequencies plays an important role in determining system behavior. In unimodal distributions, continuous phase transitions occur due to external forcing \cite{MSmall_2025}.
Bimodal frequency distributions split a system into two groups and are seen in real-world systems such as power grids (generators and consumers), breast cancer data where tumors are either ER– or ER+, and human communication patterns.
This division leads to more complex dynamics, including time-dependent traveling and standing wave states \cite{Simona_2018_bi}. Although previous studies have examined inertia and bimodal distributions \cite{gao2021synchronized}, they have not explored their combined effect in the presence of external forcing.

In this work, we investigate the synchronization transition in the Kuramoto model with inertia under external forcing, focusing on bimodal frequency distributions. We analyze how variations in forcing strength and inertia together influence the system, and characterize the resulting dynamical states. We discover that as the forcing strength increases, the transition changes from first-order discontinuous to second-order continuous. Our analytical self-consistent equation framework captures these behaviors and provides a consistent explanation of the observed transitions.
\\

\paragraph*{\bf{Model:}}
We study the effects of external forcing on Kuramoto oscillators with inertia. The equation governing the dynamics of $N$ globally coupled oscillators is given by
\begin{equation}
m \ddot{\theta}_i
= -\dot{\theta}_i + \Omega_i
+ \frac{K}{N}\sum_{j=1}^{N} \sin(\theta_j - \theta_i)
- F \sin \theta_i .
\label{eq1}
\end{equation}
In Eq.~(\ref{eq1}), $m$ is the inertia and $K$ is the coupling strength. The quantities $\theta_i$, $\dot{\theta}_i$ and $ \ddot{\theta_i}$ denote the instantaneous phase, angular velocity and angular acceleration of the $i$th oscillator, respectively. The values $\Omega_i$ represents the intrinsic frequency of the $i$th oscillator, drawn from a symmetric bimodal Gaussian distribution $g(\Omega)$ with mean zero and standard deviation $\sigma$. $F$ is the strength of the external forcing term. We decouple the differential equations in Eq.~(\ref{eq1}) by introducing a complex order parameter:
\begin{equation}
z = r e^{i\psi} = \frac{1}{N}\sum_{j=1}^{N} e^{i\theta_j}.
\label{eq2}
\end{equation}
The magnitude $r$ of the order parameter measures phase coherence, and $\psi$ represents the average phase. When the oscillator phases are uniformly distributed on the unit circle, $r$ takes $0$ value corresponding to an incoherent state. In contrast, in a coherent state all oscillators form a cluster locked to the mean phase $\psi$, and $r = 1$. Using Eq.~(\ref{eq2}), Eq.~(\ref{eq1}) can be rewritten as
\begin{equation}
m \ddot{\theta}_i
= -\dot{\theta}_i + \Omega_i
+ K r \sin(\psi - \theta_i)
- F \sin\theta_i .
\label{eq3}
\end{equation}
Choosing the rotating frame such that $\psi=0$, the equation simplifies to
\begin{equation}
m\ddot{\theta}_i = -\dot{\theta}_i + \Omega_i - (q+F)\sin\theta_i ,
\label{eq4}
\end{equation}
where $q = K r$.
\\

\paragraph*{\bf{Numerical Results:}}
For the simulation protocol, we decompose Eq.~(\ref{eq3}) into a pair of first-order differential equations and solve them numerically using the Runge-Kutta 4 algorithm.
\begin{align}
\dot{\theta}_i &= \omega_i, \\
m \dot{\omega}_i &= -\omega_i + \Omega_i
+ K r \sin(\psi - \theta_i)
- F \sin\theta_i .
\end{align}
We simulate a system of $N = 10^4$ oscillators for both forward and backward processes. The intrinsic frequencies are drawn from a bimodal Gaussian distribution given by 
\[
g(\Omega) = \frac{1}{2\sqrt{2\pi}\sigma}
\left[
\exp\!\left(-\frac{(\Omega - \Omega_0)^2}{2\sigma^2}\right)
+
\exp\!\left(-\frac{(\Omega + \Omega_0)^2}{2\sigma^2}\right)
\right]
\] with $\Omega_0 = 1.5$ and $\sigma = 0.5$.
For fixed values of $m$ and $F$, the forward process starts with random initial phases $\theta \in [0, 2\pi)$ with initial velocities $\dot{\theta} = 0$. The coupling strength $K$ is increased adiabatically from $0$ to $10$ in steps of $\Delta K = 0.01$. At each step, the system evolves for a sufficiently long time, and the final state is used as the initial condition for the next value of $K$. This ensures a continuous evolution of the system along the forward branch. In the backward process, $K$ decreases from $10$ to $0$, starting from a coherent initial state. As in the forward case, the final state at each step serves as the initial condition for the next, allowing us to capture possible hysteresis effects. In all cases, we discard an initial transient period to allow the system to reach a steady state. The order parameter is then computed by averaging over time in the steady-state regime.

In the absence of forcing, $F = 0$ (Fig. 1a), the forward branch starts from an incoherent state ($r=0$). Around $K = 2$, a discontinuous jump occurs, and the system reaches a partially coherent state (or a standing wave state). A further increase in $K$ causes the system to reach the coherent state via another discontinuous jump. In the backward branch, the system starts from a coherent state and, as $K$ decreases, it passes through a partially coherent state before reaching to an incoherent state.

In the presence of forcing, $F = 1$ (Fig. 1b), the forward branch starts from a finite $r$ value. A discontinuous increase in $r$ occurs as $K$ increases, and the system reaches a partially coherent state. Further increasing $K$ eventually takes the system to a coherent state at a lower $K$ value than in the $F = 0$ case (Fig. 1a). In the backward branch, the system starts from a coherent state and reaches an incoherent state directly via a discontinuous transition. A detailed discussion of the underlying mechanism is presented in a later section.
\\

\paragraph*{\bf{Analytical Treatment:}}
In the thermodynamic limit $N\rightarrow\infty$, the order parameter is written as
\[
r e^{i\psi}
=
\int_{-\infty}^{\infty}
\int_{-\pi}^{\pi}
e^{i\theta}\rho(\theta,\Omega)g(\Omega)d\theta d\Omega .
\]
Here, $\rho(\theta,\Omega)\,d\theta$ represents the fraction of oscillators in the continuum with intrinsic frequency $\Omega$, whose phase lies between $\theta$ and $\theta + d\theta$ on the unit circle.

In the steady state, the population of oscillators can be divided into two groups according to their intrinsic frequencies. One group synchronizes with the mean phase and becomes phase-locked, while the other group consists of drifting oscillators that move continuously relative to the locked ones. Consequently, the overall phase coherence, denoted by $r$, can be written as the sum of the contributions from the locked oscillators $(r^l)$ and the drifting oscillators $(r^d)$, such that $r = r^l + r^d$. By introducing the rescaled time variable $\tau = \sqrt{\frac{q+F}{m}}t$, Eq.~(\ref{eq4}) in the continuum limit can be rewritten as the following second-order differential equation
\begin{equation}
\ddot{\theta} = -\alpha \dot{\theta} + \beta - \sin\theta ,
\label{eq5}
\end{equation}

\begin{figure}[htbp]
  \includegraphics[width=0.8\linewidth]{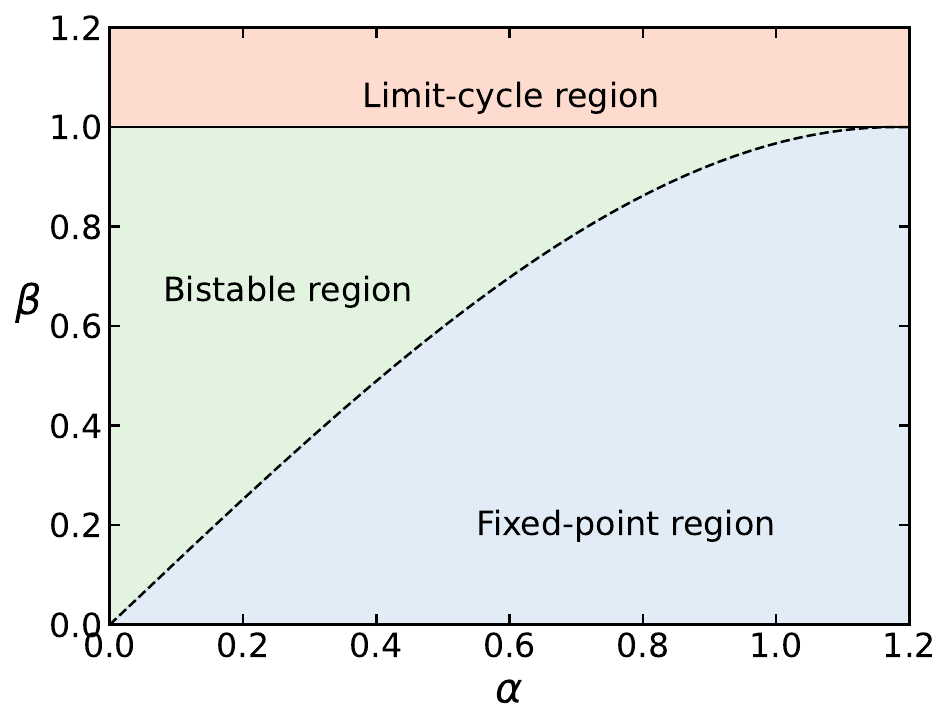}
  \caption{Phase diagram in the $\alpha$--$\beta$ parameter space for Eq.~(\ref{eq5}), where $\alpha = 1/\sqrt{(q+F)m}$ and $\beta = \Omega/(q+F)$.}
  \label{fig:alpha-beta}
\end{figure}
where $\alpha = \frac{1}{\sqrt{(q+F)m}}$ and $\beta = \frac{\Omega}{q+F}$.

For $\beta < 1$, the system admits two fixed-point solutions obtained by setting $\dot{\theta}=0$ and $\ddot{\theta}=0$. These correspond to a saddle point and a stable node \cite{Strogatz2014}. At $\beta = 1$, the system undergoes a saddle-node bifurcation, where the two fixed points collide and annihilate, yielding a unique stable limit-cycle solution for $\beta > 1$. Thus, as illustrated in Fig.~\ref{fig:alpha-beta} the dynamics can be classified into three regimes: (i)  limit-cycle regime for $\beta > 1$, (ii) bistable regime approximately defined by $\frac{4\alpha}{\pi} < \beta \le 1$, and (iii) fixed-point regime roughly characterized by $\beta \le \frac{4\alpha}{\pi}$.

Following Ref.~\cite{Tanaka1997_PD, Tanaka1997_PRL}, instead of analyzing the system in full generality, we perform the self-consistency analysis by separating the dynamics into forward ($f$) and backward ($b$) processes. In the forward process, the system starts from an incoherent state and $K=0$. Under this condition, the parameters $\alpha$ and $\beta$ take large values, placing the oscillators in the limit-cycle regime. As $K$ is increased adiabatically, the oscillators remain in the basin of attraction of the stable limit cycle even after crossing $\beta = 1$ ($\Omega = q$). They eventually join the locked cluster only when $\beta \approx \frac{4}{\pi}\alpha$ ($\Omega \approx \frac{4}{\pi}  \sqrt{\frac{q + F}{m}}$), below which the limit cycle disappears. In contrast, in the backward process we begin with a large value of $K$, where the oscillators lie in the fixed-point regime. In this regime, all the oscillators are phase-locked, forming a coherent cluster \textbf{($r \approx 1$)}. As $K$ is decreased adiabatically, the oscillators remain in the basin of attraction of the stable node until $\beta = 1$, where the stable node vanishes via a saddle-node bifurcation. 
Consequently, in the backward process oscillators with $|\Omega| \le q + F = \Omega_b$ contribute to the locked population. In contrast, during the forward process only oscillators with $|\Omega| \le \frac{4}{\pi} \sqrt{\frac{q + F}{m}} = \Omega_f$ remain locked, while those with $\Omega > \Omega_f$ drift around the locked cluster.

For  $|\Omega|\le\Omega_{f,b}$, the contribution of the locked oscillators to the overall coherence is given by $r^{l}
=
\int_{-\Omega_{f,b}}^{\Omega_{f,b}}
e^{i\sin^{-1}\!\left(\frac{\Omega}{q+F}\right)} g(\Omega)d\Omega .$
Since the frequency distribution is symmetric, $g(-\Omega)=g(\Omega)$, the imaginary part vanishes. Taking only the real part, and $\theta_{f,b}=\sin^{-1}\!\left(\frac{\Omega_{f,b}}{q+F}\right)$ we obtain
\begin{equation}
     r^{l}
=
(q+F)
\int_{-\theta_{f,b}}^{\theta_{f,b}}
\cos^2\theta\,g[(q+F)\sin\theta]\,d\theta .
\label{eq6}
\end{equation}
The contribution to the overall coherence from the drifting oscillators can be obtained as $r^d = \int_{|\Omega|>\Omega_{f,b}} \int_{-\pi}^{\pi} e^{i\theta}\,\rho_d(\theta,\Omega)\,g(\Omega)\, d\theta\, d\Omega ,$ where $\rho_d(\theta,\Omega)$ denotes the density of drifting oscillators. For drifting oscillators the density satisfies $\rho_d(\theta,\Omega) \propto 1/|\dot{\theta}|$~\cite{Tanaka1997_PRL}. The normalization condition for $\rho_d(\theta,\Omega)$ is $\int_{-\pi}^{\pi} \rho_d(\theta,\Omega)\, d\theta
=
\int_{0}^{T} \rho_d(\theta,\Omega)\dot{\theta}\, dt
= 1,$ for a given $\Omega$, where $T$ is the period of the limit-cycle solution. This yields $\rho_d(\theta,\Omega)=\frac{1}{|\dot{\theta}|T}.$ Substituting this expression into the formula for $r_d$ gives
\begin{equation}
    r^d =
\int_{|\Omega|>\Omega_{f,b}}
\frac{1}{T}\int_{0}^{T} e^{i\theta}\, dt \, g(\Omega)\, d\Omega .
\label{eq7}
\end{equation}
\noindent
To evaluate $r_d$, we first obtain an approximate analytic expression for the limit-cycle solution of Eq.~(\ref{eq5}). Following the approach described in Ref.~\cite{JGao_2018}, $\dot{\theta}$ is expanded as a Fourier series in $\theta$ while retaining only the first harmonic, $\dot{\theta}(\theta) = A_0 + A_1 \cos\theta + B_1 \sin\theta .$ Substituting this form into Eq.~(\ref{eq5}) yields the coefficients in terms of
$\alpha = \frac{1}{\sqrt{(q + F)m}}$ and $\beta = \frac{\Omega}{q + F}$, such that the first harmonic vanishes. This gives
\begin{equation}
    \dot{\theta}(\theta)=
\frac{\beta}{\alpha}
+
\frac{\alpha^2}{\alpha^4+\beta^2}
\left(
\frac{\beta}{\alpha}\cos\theta
-
\alpha\sin\theta
\right).
\label{eq8}
\end{equation}
The phase $\theta(t,\Omega)$ can then be obtained by integrating Eq.~(\ref{eq8}) with respect to time \cite{JGao_2018}. Noting that $\theta(t,-\Omega)=-\theta(t,\Omega)$ and $g(-\Omega)=g(\Omega)$, the imaginary contribution in Eq.~(\ref{eq7}) vanishes. Therefore,
\begin{equation}
    r^d =
\int_{|\Omega|>\Omega_{f,b}}
\langle \cos\theta \rangle\, g(\Omega)\, d\Omega .
\label{eq9}
\end{equation}
\noindent
The quantity $\langle \cos\theta \rangle$ can be evaluated as $\langle \cos\theta \rangle =
\frac{1}{T}\int_{0}^{T}\cos\theta\, dt
=
\frac{\int_{0}^{2\pi} \frac{\cos\theta}{\dot{\theta}}\, d\theta}
{\int_{0}^{2\pi} \frac{1}{\dot{\theta}}\, d\theta},$ which yields
\[
\langle \cos\theta \rangle = 
\frac{\beta}{\alpha}
\left[ \sqrt{
\frac{\beta^2}{\alpha^2}
-
\frac{\alpha^2}{\beta^2+\alpha^4}}
-
\frac{\beta}{\alpha}
\right].
\]

Finally from Eqs.~(\ref{eq6}) and (\ref{eq9}) the self-consistent equation for the order parameter is
\begin{equation}
\begin{aligned}
r ={}& 2(q+F)\int_{0}^{\theta_{f,b}}
\cos^2\theta \, g[(q+F)\sin\theta] \, d\theta \\
&+ 2\int_{\Omega_{f,b}}^{\infty}
\langle \cos\theta \rangle \, g(\Omega)\, d\Omega .
\label{eq10}
\end{aligned}
\end{equation}
From Eq.~(\ref{eq10}), we can understand the steady-state behavior of the system under external forcing for any given distribution of intrinsic frequencies.
For bimodal frequency distribution, the self-consistent equation, Eq.~(\ref{eq10}), is solved numerically to determine the nontrivial branch of the order parameter in the backward process, as illustrated in Fig.~\ref{fig:rk_f0_f1}.
We compare the numerical results with the analytical solution obtained from the self-consistent Eq.~(\ref{eq10}). The analytical behavior of $r$ agrees well with the numerical results in the backward branch. This agreement confirms that external forcing keeps a fraction of oscillators locked and delays the loss of coherence in the system.
\\

\paragraph*{\bf{Origin of the suppression of the intermediate state:}}
For a fixed forcing strength $F = 1$ and inertia $m = 1$, Fig.~[\ref{fig:rk_f0_f1}(b)] illustrates how the order parameter $r$ evolves as $K$ is varied during both the forward and backward continuation processes.
\begin{figure}[t]
\includegraphics[width=0.48\textwidth]{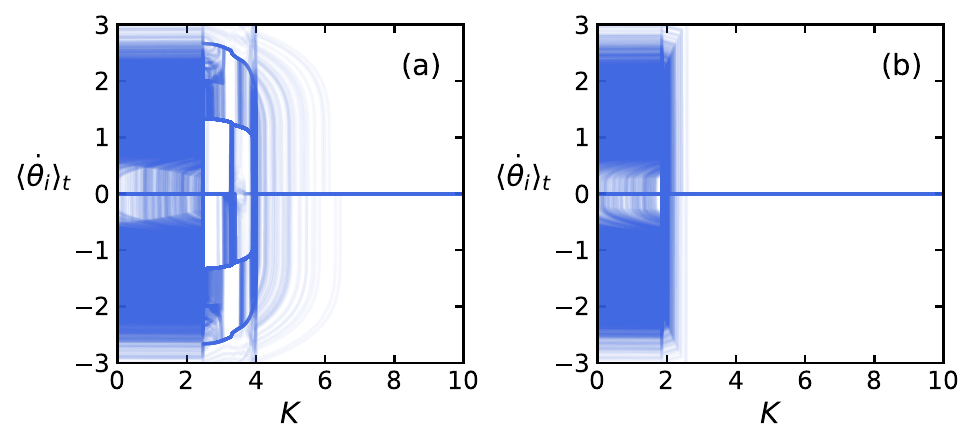}
\caption{
Formation of synchronized clusters in the mean velocities $\langle \dot{\theta}_i \rangle_t$ as a function of $K$ for the forward (a) and backward (b) branches at $F = 1$ and $m = 1$.
}
\label{fig:velocities}
\end{figure}
In the forward process, the system initially (for $K=0$) exhibits a finite value of $r$, arising from the frequency entrainment of oscillators arising due the external forcing  (Fig.~\ref{fig:velocities}(a)) as also observed for unimodal distribution \cite{Inertia_forcing_2023}. 
As $K$ increases, oscillators begin to synchronize yielding a  discontinuous transition to a partially coherent state in contrast with smooth second-order transition manifested by unimodal distribution. Since the frequency distribution is bimodal, oscillators group around two peaks, forming two synchronized clusters \cite{gao2021synchronized}. Along with these primary clusters, secondary synchronized groups may also appear, while a few oscillators remain entrained to the external forcing ($\Omega_i = 0$). Because of this forcing, the system requires a slightly larger $K$ value to reach the partially coherent state, as a fraction of oscillators stays bound to the external drive. With a further increase in $K$, the oscillators that were initially entrained to $F$ begin to drift and gradually merge into two synchronized clusters. There is a temporary decrease in $r$ value before the discontinuous transition, because a few oscillators drift away from the cluster. Around $K = 4$, a second discontinuous transition in $r$ takes place, corresponding to the merging of these synchronized clusters (Fig.~\ref{fig:velocities}(a)). In the presence of external forcing, the system requires a smaller $K$ value to reach the fully coherent state, since both the mutual interactions and the forcing term act together to lock the oscillators. Beyond this point, $r$ increases continuously with increasing $K$, eventually leading the system to a fully synchronized state.

In the backward branch, the system is initialized in a fully coherent state. As $K$ decreases, $r$ initially decreases continuously. However, with a further reduction in $K$, the intrinsic dynamics begin to dominate over the forcing, causing oscillators to drift and resulting in a discontinuous drop in $r$, as shown in Fig.~\ref{fig:rk_f0_f1}(b). From Fig.~\ref{fig:velocities}(b), it is evident that all oscillators remain locked at larger values of $K$. Around $K \approx 2$, oscillators start to drift away from the synchronized cluster, leading to a sudden decrease in $r$.
In the low $K$ regime, the system remains in a partially coherent state rather than complete incoherence, as in the case of $F = 0$ (Fig.~\ref{fig:rk_f0_f1}) due to entrainment of a few oscillators to the external forcing. External forcing delays the loss of coherence by entraining the oscillators and suppressing the formation of a standing wave state in the backward branch. In contrast, the standing wave state is clearly observed in the absence of forcing (Fig.~\ref{fig:rk_f0_f1}(a)).
\\


\begin{figure}[t]
\includegraphics[width=0.48\textwidth]{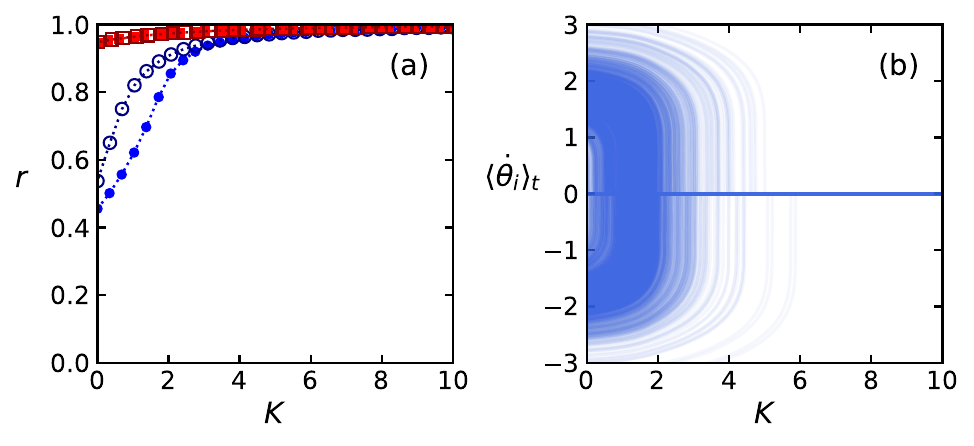}
\caption{
Effect of external forcing on hysteresis and synchronized cluster formation.
(a) $r$ as a function of $K$ for $m = 1$, with $F = 2$ (blue circles) and $F = 5$ (red squares); filled and open markers denote forward and backward branches, respectively, illustrating that the hysteresis region decreases with increasing forcing.
(b) Time-averaged velocities $\langle \dot{\theta}_i \rangle_t$ as a function of $K$, showing the formation of synchronized clusters in the forward branch at $F = 2$ and $m = 1$.
}
\label{fig:order_velocity}
\end{figure}

\paragraph*{\bf{Effect of forcing on hysteresis region:}}
External forcing strongly affects the hysteresis behavior of the system, i.e., the region enclosed between the forward and backward transition branches in Fig.~\ref{fig:order_velocity}(a). As explained earlier, forcing changes the conditions under which oscillators lock in both the forward and backward branches.
In the case of moderate forcing strength ($F = 2$), the forward branch begins from a partially coherent state, with a few oscillators entrained to the external drive (Fig.~\ref{fig:order_velocity}(b)). As $K$ increases, more oscillators join the cluster, leading to a continuous increase in coherence. In the backward branch, all oscillators start from a fully coherent state. As $K$ decreases, their phases spread out, leading to a gradual loss of coherence. 
\begin{figure}[t]
\includegraphics[width=0.4\textwidth]{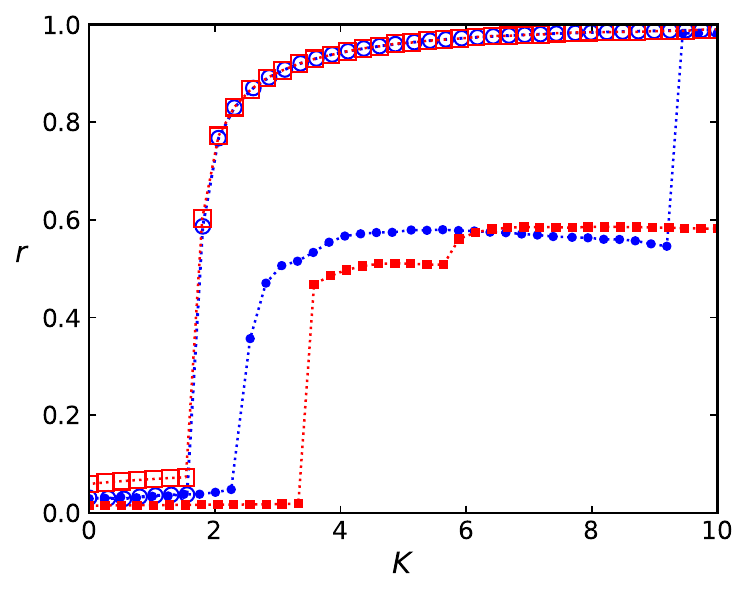}
\caption{
Effect of inertia on synchronization and hysteresis, illustrating that increasing $m$ delays the onset of synchronization and enhances the hysteresis region.
$r$ as a function of $K$ at $F = 1$, where blue circles correspond to $m = 3$ and red squares to $m = 5$; filled and open markers denote forward and backward branches, respectively.
}
\label{fig:m3_m5}
\end{figure}
For a larger forcing strength ($F = 5$), all oscillators entrain to the external forcing, resulting in a coherent state even in the absence of mutual coupling. As a result, the forward and backward branches overlap and the hysteresis region shrinks in the observed $K$ range (Fig.~\ref{fig:order_velocity}(a)).
\\

\paragraph*{\bf{Role of inertia:}}
As shown in Fig.~\ref{fig:m3_m5}, for $m = 3$, the system on the forward branch initially enters a partially coherent state, as discussed earlier. As $K$ increases, the system approaches a partially coherent state, and further increases in $K$ lead to a coherent state.
For $m = 5$, the value of $r$ in the forward branch decreases slightly compared to $m = 3$. Since $m$ changes the locking condition, the range of $K$ over which the partially coherent state remains stable extends. Therefore, a greater $K$ value is required for the system to achieve full coherence. This adjustment shifts the forward transition point and increases the hysteresis region. In contrast, the backward branch for both cases overlaps, as it remains largely unchanged and is not affected by inertia.

To understand the role of forcing and inertia more clearly, we consider the single oscillator limit by setting $K=0$. For a single oscillator, we get 
\begin{equation}
m \ddot{\theta} = -\dot{\theta} + \Omega - F \sin{\theta}.
\label{eq19}
\end{equation}
For the locked state, $\dot{\theta} = 0$ and $\ddot{\theta} = 0$, yield two fixed points, one stable and the other a saddle, given by
\[
\theta^* = \sin^{-1}\!\left(\frac{\Omega}{F}\right), 
\quad \text{and} \quad 
\theta^* = \pi - \sin^{-1}\!\left(\frac{\Omega}{F}\right).
\] 
For fixed points to exist, the condition $|\Omega| \leq F$ must be satisfied; otherwise, Eq.~(\ref{eq19}) admits only a limit-cycle solution. This condition provides an analytical form for the existence of fixed points, but it does not fully capture the system's dynamical behavior. To explore this further, we numerically simulate Eq.~(\ref{eq19}) for a fixed value of $F$ and different values of $m$, with $\Omega$ varying from $-3$ to $3$. For each value of $\Omega$, we use random initial conditions and compute the time-averaged angular velocity after discarding the initial transient.

\begin{figure}[t]
\includegraphics[width=\linewidth]{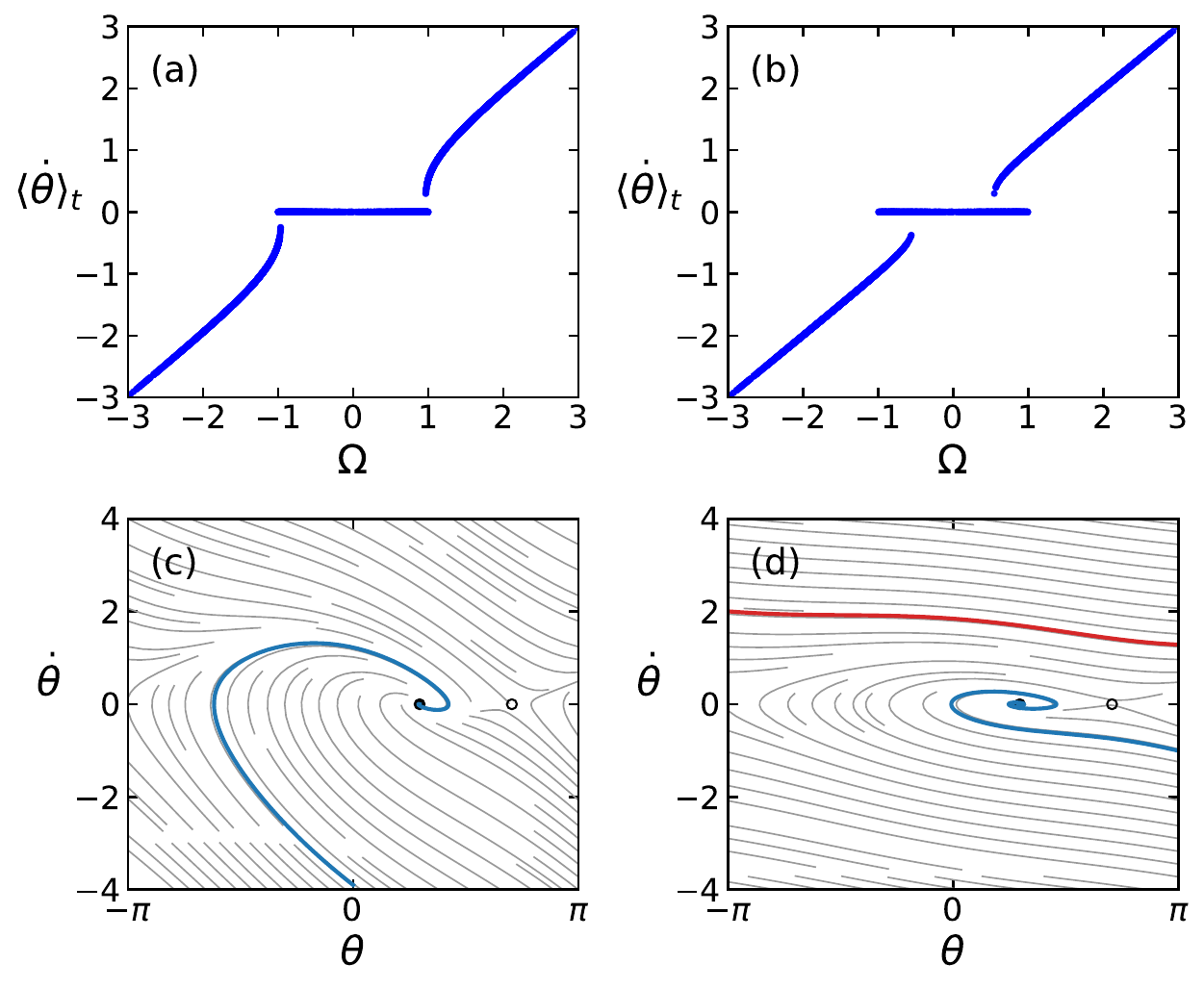}
\caption{
Role of inertia in bistability and phase-space structure. Top panels: Time-averaged velocity $\langle \dot{\theta} \rangle_t$ as a function of intrinsic frequency $\Omega$ for $K = 0$ and $F = 1$, with panels (a) and (b) corresponding to $m = 1$ and $m = 5$, respectively.
Bottom panels: Phase portraits in the $(\theta, \dot{\theta})$ plane for Eq.~(\ref{eq19}), with panels (c) and (d) corresponding to $m = 1$ and $m = 5$ at $\Omega = 0.8$. Filled (open) circles denote stable (saddle) fixed points, while the blue and red curves represent the fixed-point trajectory and the limit-cycle trajectory, respectively.
}
\label{fig:combined}
\end{figure}

As depicted in Fig.~\ref{fig:combined}(a), in the small inertia limit ($m = 1$), all values of $\Omega$ satisfying the locking condition are entrained to the forcing. For moderate inertia ($m = 5$, Fig.~\ref{fig:combined}(b)), a bistable region emerges over a range of $\Omega$, where the long-term dynamics depend on the initial conditions. With a further increase in $m$, this bistable region widens, demonstrating that inertia enables the coexistence of locked and drifting states over a broader range of $\Omega$.

As reflected by Figs.~\ref{fig:combined} (c) and (d), increasing $m$ leads to a reduction in the basin of attraction of the fixed point. For large $m$, only those trajectories that start within the basin of the fixed point tend to converge to the fixed point. In contrast, trajectories that start outside this reduced basin tend to drift and eventually settle onto the limit-cycle state. As $m$ increases, the basin of attraction of the fixed point shrinks further, and bistability becomes more pronounced due to the coexistence of fixed point and limit-cycle solutions. As a result, in the forward branch, fewer oscillators can reach the fixed point at lower $K$ values, leading to a smaller initial value of $r$ for larger $m$ value than in the smaller $m$ values.
\\

\begin{table*}[t]
\begin{tabular}{c p{5cm} p{10cm}}
\hline
\textbf{No.} & \textbf{Model} & \textbf{Key Phenomena} \\
\hline

1 & KMI + unimodal 
& First-order phase transition and hysteresis \cite{Tanaka1997_PRL} \\

2 & KMI + unimodal + forcing 
& Continuous synchronization transition \cite{MSmall_2025} \\

3 & KMI + bimodal 
& Standing wave and traveling wave states \cite{acebron2000synchronization} \\

4 & KMI + bimodal + forcing 
& Suppression of standing wave states due to interplay between intrinsic frequencies and external forcing; discontinuous backward transition \\

\hline
\end{tabular}
\caption{Summary of key results in the Kuramoto model with inertia (KMI) for unimodal and bimodal frequency distributions, with and without external forcing. The first three rows list established results, while the fourth row presents  result of Eq.~(\ref{eq1}).}
\label{tab:results}
\end{table*}

\paragraph*{\bf{Bi-delta Distribution:}}

\begin{figure}[t]
\includegraphics[width=\linewidth]{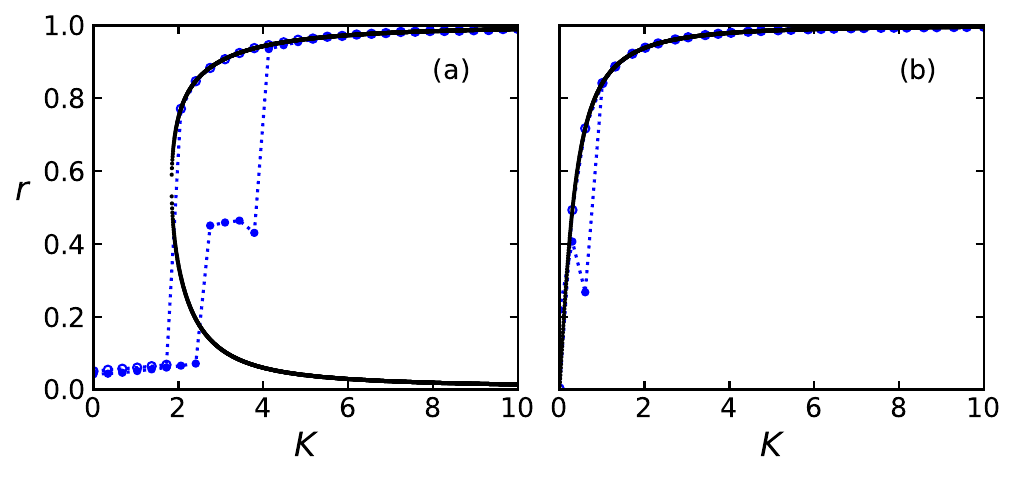}
\caption{
Comparison of synchronization for bimodal Gaussian (a) and bi-delta (b) distributions at $F=1$, $m=1$. 
Blue filled (open) circles denote the forward (backward) branch, and the black line represents the analytical solutions obtained from Eqs.~(\ref{eq10}) and~(\ref{eq14}) for (a) and (b), respectively.
}
\label{fig:gaussian_bidelta_comparison}
\end{figure}

We numerically solved the self-consistent Eq.~(\ref{eq10}) for a bimodal Gaussian distribution (Fig.~\ref{fig:gaussian_bidelta_comparison}(a)). To obtain a closed form expression, we then considered a bi-delta distribution \cite{montbrio2006time}, $g(\Omega) = \tfrac{1}{2}\left[\delta(\Omega-\Omega_0)+\delta(\Omega+\Omega_0)\right]$ with $\Omega_0 = 1$. 
For $q+F \ge 1$, both frequency peaks lie in the locked region and the drifting contribution vanishes. Substituting the distribution into Eq.~(\ref{eq10}) gives
\begin{align*}
r = (q+F)\int_{0}^{\pi/2} \cos^2\theta 
\Big[\delta\!\big((q+F)\sin\theta - 1\big) \\
+ \delta\!\big((q+F)\sin\theta + 1\big)\Big] d\theta.
\end{align*}
Only the solution $(q+F)\sin\theta^* = 1$ contributes in $[0,\pi/2]$, which yields $r = \cos\theta^*$ with $\sin\theta^* = (q+F)^{-1}$. Therefore, for any forcing strength ($F$), the self-consistent solution of the backward branch is found to be 
\begin{equation}
r = \frac{\sqrt{(q+F)^2 - 1}}{q+F}, \quad q = Kr.
\label{eq14}
\end{equation}
The Eq.~(\ref{eq14}) indicates that, for ($F \geq 1$), the oscillators remain locked across the entire range of $K$ considered. In the presence of forcing, the backward branch exhibits a smooth, continuous desynchronization transition
(analytical result depicted as the black line in Fig.~\ref{fig:gaussian_bidelta_comparison}(b)), in contrast to the discontinuous behavior observed for the bimodal Gaussian distribution (Fig.~\ref{fig:gaussian_bidelta_comparison}(a)). However, in both distributions, the intermediate standing wave states are suppressed, as external forcing entrains the oscillators, and the system loses coherence at a smaller value of $K$ in the bi-delta case. The analytical predictions agree well with the numerical results in the backward branch. In the forward branch, the two-cluster state emerges immediately for $K > 0$, and coherence is achieved at a lower $K$. In contrast, for the bimodal Gaussian case, the wider frequency spread delays cluster formation, requiring a comparatively larger $K$ to reach a coherent state.
\\

\paragraph*{\bf{Conclusions:}}
In this work, we investigated the synchronization transition in the Kuramoto model with inertia under external forcing for a bimodal frequency distribution. First, using numerical simulations, we examined how the dynamical behavior evolves with coupling strength and how varying the forcing strength influences the collective dynamics. We then derived the corresponding analytical self-consistent equation for the system under forcing. These findings are summarized in Table~\ref{tab:results}, which highlights the key differences between unimodal and bimodal distributions, both in the absence and presence of forcing.
We find that for small forcing, the backward transition becomes discontinuous rather than following the usual route to incoherence via a standing wave or a partially synchronized state. In contrast, for large forcing, the transition becomes continuous. Overall, external forcing significantly modifies the synchronization transition: (i) reduces the hysteresis region as the forcing strength increases, (ii) influences the stability of different dynamical states, and (iii) alters the pathway through which the system transitions between coherent and incoherent behavior. 

We study the role of inertia by analyzing its effect on the basin of attraction of the fixed point in the limiting case of a single oscillator ($K = 0$). We determine the condition for the existence of a fixed point ($|\Omega| \leq F$). Numerical results show that for small $m$ value, all trajectories in phase space converge to the fixed point. However, as $m$ increases, a bistable region emerges in which a limit-cycle solution coexists with the fixed point. Consequently, the basin of attraction of the fixed point shrinks, and the long-term dynamics become dependent on initial conditions. As a result, increasing $m$ reduces the value of $r$ in the forward branch. Moreover, since the locking condition in the forward branch depends on $m$ $\left(\Omega_f = \frac{4}{\pi} \sqrt{\frac{q + F}{m}}\right)$, increasing $m$ delays the onset of synchronization.
Overall, this indicates that in the forward branch, inertia competes with the external forcing, reduces locking, and suppresses synchronization.

Finally, we derive a closed form solution of the self-consistent equation for the backward branch by considering a bi-delta frequency distribution. The analytical expression explains the mechanism of the continuous desynchronization transition in the bi-delta distribution, in contrast to the discontinuous transition in the bimodal Gaussian distribution. However, the phenomenon of the suppression of standing wave states remains the characteristics of both the distributions.

A natural extension of the present work would be to incorporate higher-order interactions \cite{battiston2021physics} and network structure within the self-consistent framework, which may further influence the collective dynamics and synchronization transitions. Extending the analysis to time-dependent external forcing \cite{Sakaguchi_1988} in the presence of inertia remains a challenging direction, as the self-consistent approach employed here is inherently restricted to steady states \cite{gao2021synchronized}. Such extensions will offer deeper insight into synchronization in complex systems.

\section*{References}
\bibliographystyle{apsrev4-1}
\bibliography{references}

\end{document}